\def\year{2023}\relax
\documentclass[letterpaper]{article} 
\usepackage{aaai23}  
\usepackage{times}  
\usepackage{helvet}  
\usepackage{courier}  
\usepackage[hyphens]{url}  
\usepackage{graphicx} 
\usepackage{natbib}  
\usepackage{caption} 
\usepackage{bm}
\usepackage{subfigure}
\usepackage{multirow}
\usepackage{amsmath}
\usepackage{amsfonts}
\usepackage{comment}
\usepackage{color}

\DeclareCaptionStyle{ruled}{labelfont=normalfont,labelsep=colon,strut=off} 
\usepackage{algorithm}
\usepackage{algorithmic}

\usepackage{newfloat}
\usepackage{listings}

\begin{document}

\lstset{%
	basicstyle={\footnotesize\ttfamily},
	numbers=left,numberstyle=\footnotesize,xleftmargin=2em,
	aboveskip=0pt,belowskip=0pt,%
	showstringspaces=false,tabsize=2,breaklines=true}
\floatstyle{ruled}
\newfloat{listing}{tb}{lst}{}
\floatname{listing}{Listing}
%
%
\pdfinfo{
/Title (AAAI Press Formatting Instructions for Authors Using LaTeX -- A Guide)
/Author (AAAI Press Staff, Pater Patel Schneider, Sunil Issar, J. Scott Penberthy, George Ferguson, Hans Guesgen, Francisco Cruz, Marc Pujol-Gonzalez)
/TemplateVersion (2022.1)
}

\setcounter{secnumdepth}{0} 

%


\title{Long-tail Cross Modal Hashing}
\author{Paper ID: 7945}

\author{
    Zijun Gao\textsuperscript{\rm 1},
    Jun Wang\textsuperscript{\rm 2,*},
    Guoxian Yu\textsuperscript{\rm 1,2},
    Zhongmin Yan\textsuperscript{\rm 1},
    Carlotta Domeniconi\textsuperscript{\rm 3},
    Jinglin Zhang\textsuperscript{\rm 4}
}
\affiliations{
    \textsuperscript{\rm 1}School of Software, Shandong University, Jinan, China\\
    \textsuperscript{\rm 2}SDU-NTU Joint Centre for AI Research, Shandong University, Jinan, China\\
    \textsuperscript{\rm 3}Department of Computer Science, George Mason University, Fairfax, VA, USA\\
    \textsuperscript{\rm 4}School of Control Science and Engineering, Shandong University, Jinan, China\\
    zjgao@mail.sdu.edu.cn, \{kingjun, gxyu, yzm\}@sdu.edu.cn, carlotta@cs.gmu.edu, jinglin.zhang@sdu.edu.cn
}
\maketitle
\begin{abstract}
Existing Cross Modal Hashing (CMH) methods  are mainly designed for balanced data, while imbalanced data with long-tail distribution is more general in real-world. Several long-tail hashing methods have been proposed but they can not adapt for multi-modal data, due to the complex interplay between labels and individuality and commonality information of multi-modal data. Furthermore, CMH methods mostly mine the commonality of multi-modal data to learn hash codes, which may override tail labels encoded by the individuality of respective modalities. In this paper, we propose LtCMH (Long-tail CMH) to handle imbalanced multi-modal data. LtCMH firstly adopts auto-encoders to mine the individuality and commonality of different modalities by minimizing the dependency between the individuality of respective modalities and by enhancing the commonality of these modalities. Then it dynamically combines the individuality and commonality with direct features extracted from respective modalities to create meta features that enrich the representation of tail labels, and  binaries meta features to generate hash codes. LtCMH significantly outperforms state-of-the-art baselines on long-tail datasets and holds a better (or comparable) performance on datasets with balanced labels.
\end{abstract}

\section{Introduction}
Hashing aims to map high-dimensional data into a series of low-dimensional binary codes while preserving the data proximity in the original space. The binary codes can be computed in a constant time and economically stored, which meets the need of large-scale data retrieval. In real-world applications, we often want to retrieval data from multiple modalities. For example, when we input key words to information retrieval systems, we expect the systems can efficiently find out related news/images/videos from database. Hence, many cross modal hashing (CMH) methods have been proposed to deal with such tasks \cite{wang2016comprehensive,jiang2017deep,yu2022FlexCMH}.

Most CMH methods aim to find a low-dimensional shared subspace to eliminate the modality heterogeneity and to quantify the similarity between samples across modalities.
More advanced methods leverage extra knowledge (i.e., labels, manifold structure, neighbors coherence) \cite{jiang2017deep, yu2021deep, yu2022FlexCMH} to more faithfully capture the proximity between multi-modality data to induce hash codes. However, almost all of them are trained and tested on hand-crafted balanced datasets, while \emph{imbalanced} datasets with long tail labels are more prevalence in real-world. Some recent studies \cite{chen2021long,cui2019class} have reported that the imbalanced nature of real-world data greatly compromises the retrieval performance.

Real-world samples typically have a skewed distribution with \textbf{long-tail labels}, which means that a few labels (a.k.a. \emph{head labels}) annotate to many samples, while the other labels (a.k.a. \emph{tail labels}) have a large quantity but each of them is only annotated to several samples \cite{liu2020deep,chen2021long}. 
It is a challenging task to train a general model from such distribution, because the head labels gain most of the attention while the tail ones are underestimated. 
Another inevitable problem of long-tail hashing is the ambiguity of the generated binary codes. Due to the dimension reduction, information loss is unavoidable. In this case, hash codes learned from data-poor tail labels more lack the discrimination, which seriously confuses the retrieval results. 

\begin{figure}[ht!bp]
\centering
\includegraphics[width=8.5cm]{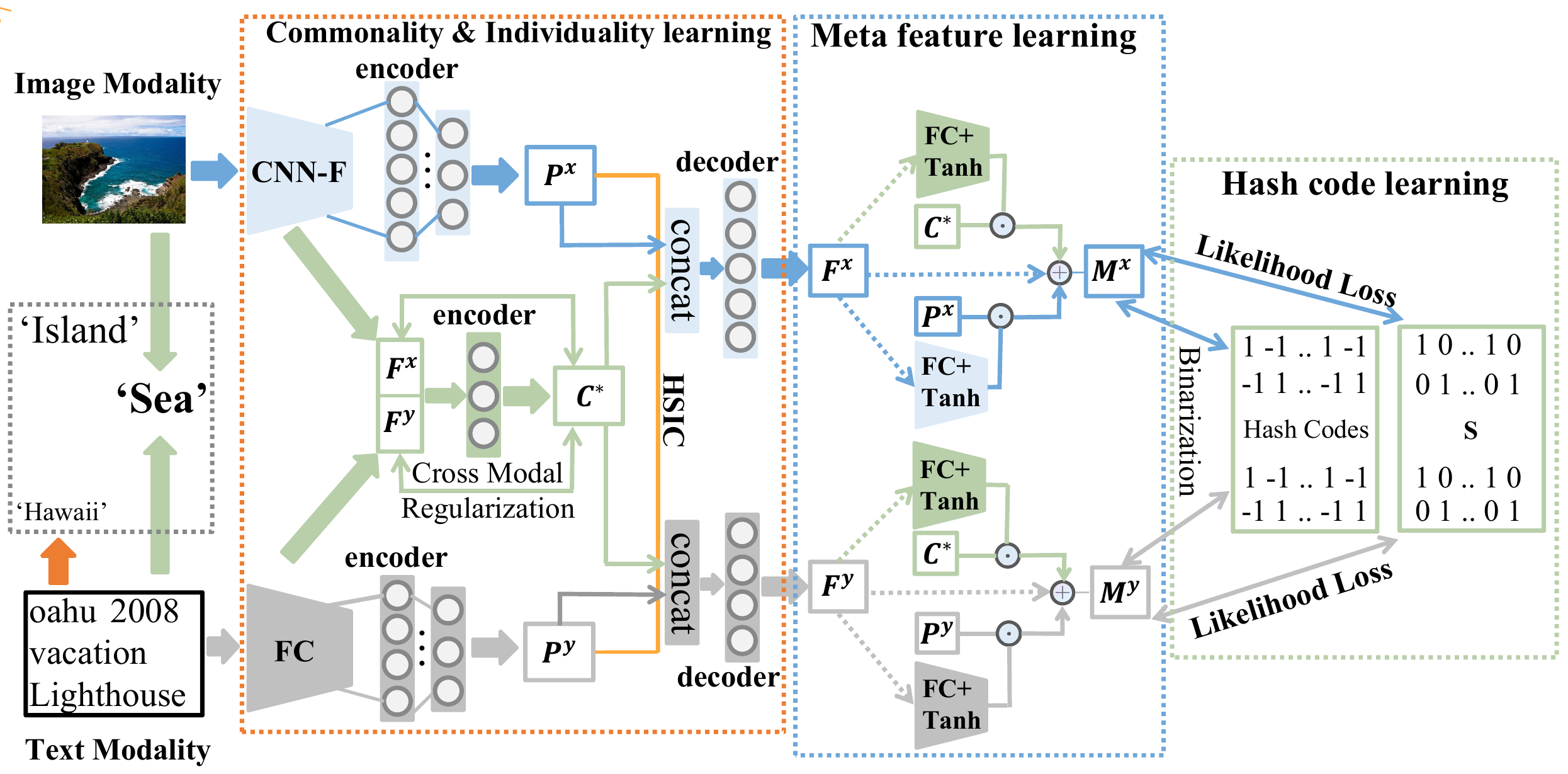}
\caption{The schematic framework of LtCMH. {The direct image/text features $\mathbf{F}^x$/$\mathbf{F}^y$ are first extracted by CNN-F and  FC (fully-connected) networks with two layers. LtCMH then uses different auto-encoders to mine the commonality $\mathbf{C}^*$ and individuality $\mathbf{P}^x$/$\mathbf{P}^y$ of multi-modal data via cross-modal regularization and  Hilbert-Schmidt independence criterion (HSIC). After that, it creates meta features $\mathbf{M}^x$/$\mathbf{M}^y$ by fusing $\mathbf{C}^*$, $\mathbf{P}^x$/$\mathbf{P}^y$ and $\mathbf{F}^x$/$\mathbf{F}^y$. Next, LtCMH binarizes meta features into hash codes. The head label `Sea' can be consolidated from the commonality of image and text modality, while tail label `Hawaii' can only be obtained from the text-modality. The complex relations between labels and heterogeneous modalities can be more well explored by the commonality and individuality. }}
\label{fig:framework}
\end{figure}

Several efforts have been made toward long-tail single-modal hashing via knowledge transfer \cite{liu2020deep} and information augmentation \cite{chu2020feature,wang2020devil,kouattention}. LEAP \cite{liu2020deep} augments each instance of tail labels with certain disturbances in the deep representation space to provide higher intra-class variation to tail labels. OLTR \cite{liu2019large} and LTHNet \cite{chen2021long} propose a meta embedding module to transfer knowledge from head labels to tail ones. 
{OLTR uses cross-entropy classification loss, which doesn't fully match the retrieval task. In addition, these  long-tail hashing methods cannot be directly adapted for multi-modal data, due to the complex interplay between long-tail labels and heterogeneous data modalities. The head labels can be easily represented by the commonality of multiple modalities, but the tail labels are often represented by individuality of a particular modality. For example, in the left part of Figure \ref{fig:framework}, tail label `Hawaii' can only be read out from the text-modality. On the other hand, head label `Sea' can be consolidated from the commonality of image and text modality and head label ’Island’ comes from the image modality alone. From this example, we can conclude that both the \textbf{individuality} and \textbf{commonality} should be used for effective CMH on long-tail data, which also account for the complex interplay between head/tail labels and multi-modal data. Unfortunately, most CMH methods solely mine the commonality (shared subspace) to learn hashing functions and assume balanced multi-modal data \cite{wang2016comprehensive,jiang2017deep,yu2022FlexCMH}. 
}


To address these problems, we propose LtCMH (Long-tail CMH) to achieve CMH on imbalanced multi-modal data, as outlined in Figure \ref{fig:framework}.
Specifically, we adopt different auto-encoders to mine individuality and commonality information from multi-modal data in a collaborative way. We further use the Hilbert-Schmidt independence criterion (HSIC) \cite{gretton2005measuring} to extract and enhance the individuality of each modality, and cross-modal regularization to boost the commonality. We model the interplay between head/tail labels and multi-modal data by meta features dynamically fused from the commonality, individuality and direct features extracted from respective modalities. The meta features can enrich tail labels and preserve the correlations between different modalities for more discriminate hash codes. We finally binarize the meta features to generate hash codes. The contributions of this work are summarized as follows:\\
(i) We study CMH on long-tail multi-modal data, which is a novel, practical and difficult but understudied topic. We propose LtCMH to achieve effective hashing on both long-tail and balanced datasets.\\
(ii) LtCMH can mine the individuality and commonality information of multi-modal data to more comprehensively model the complex interplay between head/tail labels and  heterogeneous data modalities. It further defines a dynamic meta feature learning module to enrich labels and to induce discriminate hash codes.\\
\noindent (iii) Experimental results show the superior robustness and performance of LtCMH to competitive CMH methods \cite{wang2020joint,li2018self,yu2021deep} on benchmark datasets, especially for long-tail ones.

\section{Related Work}
\subsection{Cross Modal Hashing}
Based on using the semantic labels or not, existing CMH can be divided into two types: unsupervised and supervised. Unsupervised CMH methods usually learn hash codes from data distributions without referring to supervised information. 
Early methods, such as cross view hashing (CVH) \cite{kumar2011learning} and collective matrix factorization hashing (CMFH) \cite{ding2014collective}, typically build on canonical correlation analysis (CCA).
UDCMH \cite{wu2018unsupervised} keeps the maximum structural similarity between the deep hash codes and original data. {DJSRH \cite{su2019deep} learns a joint semantic affinity matrix to reconstruct the semantic affinity between deep features and hash codes.}
Supervised CMH methods additionally leverage supervised information such as semantic labels and often gain a better performance than unsupervised ones. 
For example, DCMH \cite{jiang2017deep} optimizes the joint loss function to maintain the label similarity between unified hash codes and feature representations of each modality. BiNCMH \cite{sun2022deep} uses bi-direction relation reasoning to mine unrelated semantics to enhance the similarity between features and hash codes. 

Although these CMH methods perform well on balanced datasets, they are quite fragile on long-tail datasets, in which samples annotated with head labels gain more attention than those with tail labels  when jointly optimizing correlation between modalities \cite{zhou2020bbn}. {Given that, we try to enrich the representation of tail labels by mining the individuality and commonality of multi-modal data.  This enriched representation can more well model the relation between head/tail labels and multi-modal data.
}

\subsection{ Long-Tail Data Learning}
Long-tail problem is prevalence in real-world data mining tasks and thus has been extensively studied \cite{chawla2009data}. There are three typical solutions: re-sampling, re-weighting and transfer-learning. Re-sampling based solutions use different sampling strategies to re-balance the distribution of different labels \cite{buda2018systematic,byrd2019effect}. 
Re-weighting solutions give different weights to the head and tail labels of the loss function \cite{huang2016learning}. 
Transfer-learning based techniques learn general knowledge from the head labels, and then use these knowledge to enrich tail labels. \citet{liu2019large} proposed a meta feature embedding module to combine direct features with memory features to enrich the representation of tail labels.  \citet{liu2020deep} proposed the concept of `feature cloud' to enhance the diversity of tail ones. 

Most of these long-tail solutions manually divide samples into head labels and tail labels, which compromises the generalization ability. In contrast, our LtCMH does not need to divide head and tail labels beforehand, it leverages both individuality and commonality as memory features, rather than directly stacking direct features \cite{liu2019large,chen2021long}, to reduce information overlap. As a result, LtCMH has a stronger generalization and feature representation capability. 

\section{The Proposed Methodology}
\subsection{Problem Overview}
Without loss of generality, we firstly present LtCMH based on two modalities (image and text). LtCMH can also be applied to other data modality or extended to $\geq 3$ modalities \cite{liu2022WCHash}. Suppose $\mathcal X = \{\mathbf{x}_1, \mathbf{x}_2, \cdots, \mathbf{x}_n\}$ and $\mathcal Y= \{\mathbf{y}_1, \mathbf{y}_2, \cdots, \mathbf{y}_n\}$ are the respective image and text modality with $n$ samples, and  $\mathbf{L} = \{\mathbf{l}_1, \mathbf{l}_2,\cdots, \mathbf{l}_n\}$ is the label matrix of these samples from $c$ distinct labels. A dataset is called long-tail if the numbers of samples annotated with these $c$ labels conform to Zipf's law distribution \cite{reed2001pareto}: $z_a = z_1 \times a^{-\mu}$, where $z_a$ is the number of samples with label $a$ in decreasing order ($z_1>z_2>\cdots \gg z_c$) and {$\mu$ is the control parameter of imbalance factor (\textbf{IF} for short, IF=$z_1/z_c$).} The goal of CMH is to learn hash functions ($h^x$ and $h^y$) from $\mathcal{X}$ and $\mathcal{Y}$, and to generate hash codes via $\mathbf{b}^x=h^x(\mathbf{x})$ and $\mathbf{b}^y=h^y(\mathbf{y})$, where $\mathbf{b}^x/\mathbf{b}^y \in \{0,1\}^k$ is the length-$k$ binary hash code vector. For ease presentation, we denote $v$ as any modality, $x/y$ as the image/text modality indicator.



The whole framework of LtCMH is illustrated in Figure~\ref{fig:framework}. LtCMH firstly captures the individuality and commonality of multi-modal data, and takes the individuality and commonality as the memory features. Then it {introduces individuality-commonality selectors} on memory features, along with the direct features extracted from respective modality, to create meta features, which not only enrich the representation of tail labels, but also automatically balance the head and tail labels. Finally, it binaries meta features into binary codes through {the hash learning module}. The following subsections elaborate on these steps.



\subsection{Individuality and Commonality Learning}
Before learning the individuality and commonality of multi-modal data, we adopt Convolutional Neural Networks (CNN) to extract direct visual features $\mathbf{F}^x=CNN(\mathcal{X}) \in \mathbb{R}^{n \times d_x}$ and fully connected nerworks with two layers (FC) to extract textual features $\mathbf{F}^y=FC(\mathcal{Y}) \in \mathbb{R}^{n \times d_y}$. Other feature extractors can also be adopted for modality adaption. 
Due to the prevalence of samples of head labels, these direct features are more biased toward head labels and under-represent tail ones. To address this issue, long-tail single modal hashing solutions learn label prototypes to summarize visual features and utilize them to transfer knowledge from head to tail labels \cite{wei2022prototypical,tang2020long,chen2021long}. But they can not account for the complex interplay between head/tail labels and individuality and commonality of multi-modal data, as we exampled in Figure \ref{fig:framework} and discussed in the Introduction. Some attempts have already explored the individuality and commonality of multi-view data to improve the performance of multi-view learning \cite{wu2019mvml,yu2022M3AL}, and \citet{tan2021ICM2L} empirically found the classification of less frequent labels can be improved by the individuality. Inspired by these works, we advocate to mine the individuality and commonality information of multi-modal data with long-tail distribution. We then leverage the individuality and commonality to create meta features, which can enrich the representation of tail labels and model the interplay between labels and multi-modal data .




Because of the information overlap of multi-modal high-dimensional data, we want to mine the shared and specific essential information of different modalities in the low-dimensional embedding space. Auto-encoders (AE) \cite{goodfellow2016deep} is an effective technique that can map high-dimensional data into an informative low-dimensional representation space. AE can encode unlabeled/incomplete data and disregard the non-significant and noisy part, so it has been widely used to extract disentangled latent representation. Here, we propose a new structure for learning the individuality and commonality of multi-modal data. The proposed individuality-commonality AE (as shown in Figure \ref{fig:framework}) is similar to a traditional AE in both the encoding end of the input and the decoding end of the output. The new ingredients of our AE are the regularization of learning individuality and commonality, and the decoding of each modality via combining the individuality of this modality and commonality of multiple modalities.


As for the encoding part, we use view-specific encoders $\{f_{enI}^{v}\} $ to initialize the individuality information matrix of each modality as follows:
\begin{equation}
\label{eq7}
    \mathbf{P}^x = f_{enI}^{x}(\mathbf{F}^x) , \mathbf{P}^y = f_{enI}^{y}(\mathbf{F}^y)
\end{equation}
where $\mathbf{P}^x$ and $\mathbf{P}^y$ encode the individuality of image and text modality, respectively.
To learn the commonality of multi-modal data, we concatenate $\mathbf{F}^x$ and $\mathbf{F}^y$ and then input them into commonality encoder $f_{enC}$ as:
\begin{equation}
    \mathbf{C}^* = f_{enC}([\mathbf{F}^x, \mathbf{F}^y])  
    \label{eq2}
\end{equation}

Although $\mathbf{C}^*$ can encode the commonality of multi-modal data by fusing $\mathbf{F}^x$ and $\mathbf{F}^y$, it does not concretely consider the intrinsic distribution of respective modalities. Hence, we further define a cross-modal regularization to optimize $\mathbf{C}^*$ and to enhance the commonality by leveraging the shared labels of samples and data distribution within each modality as:
\begin{equation}
\begin{aligned}
    \mathcal{J}_1(\mathbf C^*) & = \frac{1}{2} ( \sum_{a,b=1}^c \begin{Vmatrix} \mathbf C_{\cdot a}^*-\mathbf C_{\cdot b}^*  \end{Vmatrix}^2 (\mathbf{R}_{ab}^x + \mathbf{R}_{ab}^y ))\\
    & = tr( \mathbf C^*((\mathbf{D}^x-\mathbf{R}^x)+ (\mathbf{D}^y-\mathbf{R}^y))(\mathbf C^*)^\mathrm{T} )
\end{aligned}
\label{eq3}
\end{equation}
Here $\mathbf C^*_{\cdot a}$ is the $a$-th column of $\mathbf C^*$, $tr(\cdot)$ denotes the matrix trace operator, $\mathbf{D}^{v}$ is a diagonal matrix with $\mathbf{D}_{aa}^{v} = \sum_{b=1}^c \mathbf{R}_{ab}^{v}$. $\mathbf{R}^{v}$ quantifies the similarity between different labels, it is defined as follows:
\begin{equation}
\begin{aligned}
    \mathbf{R}^{v}_{ab} &= e^{\frac{-\mathbf{H}^{v}_{ab}}{(\sigma^{v})^2}}\\
    \mathbf{H}^{v}_{ab} & = \frac{\sum_{\mathbf{f}_i^{v}\in \mathcal{\chi}_a^{v} }\mathop{\min}_{\mathbf{f}_j^{v}\in\chi_b^{v}}d(\mathbf{f}_i^{v},\mathbf{f}_j^{v})}{\begin{vmatrix}\mathcal{\chi}_a^{v}  \end{vmatrix} + \begin{vmatrix}\mathcal{\chi}_b^{v}  \end{vmatrix}} \\
    & + \frac{\sum_{\mathbf{f}_j^{v}\in\chi_b^{v} }\mathop{\min}_{\mathbf{f}_i^{v}\in\chi_a^{v}}d(\mathbf{f}_i^{v},\mathbf{f}_j^{v})}{\begin{vmatrix}\chi_a^{v}  \end{vmatrix} + \begin{vmatrix}\chi_b^{v}  \end{vmatrix}}
\end{aligned}
\end{equation}
where $\mathbf{H}^{v}_{ab}$ is the average Hausdorff distance between two sets of samples separately annotated with label $a$ and $b$, $d(\mathbf{f}_i^{v},\mathbf{f}_j^{v})$ is the Euclidean distance between $\mathbf{f}_i^{v}$ and $\mathbf{f}_j^{v}$. $\begin{vmatrix}\chi_a^{v}  \end{vmatrix}$ counts the number of samples annotated with $a$. $\sigma^{v}$ is set to the average of $\mathbf{H}^{v}$. 

Eq. \eqref{eq3} aims to learn the commonality of each label across modalities, it jointly considers the intrinsic distribution of samples annotated with a particular label within each modality, thus $\mathbf{C}^*$ can bridge $\mathcal{X}$ and $\mathcal{Y}$ and enable CMH.
We want to remark that other distance metrics can also be used to setup $\mathbf{H}^{v}$. Our choice of Hausdorff distance is for its intuitiveness and effectiveness on qualifying two sets of samples \cite{hausdorff2005set}. 
For the variants of Hausdorff distance, we choose the average Hausdorff distance because it considers more geometric relations between samples of two sets than the maximum and minimum Hausdorff distances \cite{zhou2012multi}. By optimizing Eq. \eqref{eq3} across modalities, we can enhance the quality of extracted commonality of multi-modal data.

$\mathbf{P}^x$ and $\mathbf{P}^y$ may be highly correlated, because they are simply obtained from the individuality auto-encoders without any contrast and collaboration, and samples from different modalities  share the same labels.  As such, the individuality of each modality cannot be well preserved and tail labels encoded by the individuality are under-represented. Here, we minimize the correlation between $\mathbf{P}^x$ and $\mathbf{P}^y$ to capture the intrinsic individuality of each modality. Meanwhile, $\mathbf{C}^*$ can include more shared information of multi-modal data. 

For this purpose, we use HSIC \cite{gretton2005measuring} to approximately quantify the correlation between $\mathbf{P}^x$ and $\mathbf{P}^y$, for its simplicity and effectiveness on measuring linear and nonlinear interdependence/correlation between two sets of variables. The correlation between $\mathbf{P}^x$ and $\mathbf{P}^y$ is approximated as:
\begin{equation}
    \begin{aligned}
     \mathcal{J}_2(\mathbf{P}^{v})=HSIC(\mathbf{P}^x,\mathbf{P}^y)  = (n-1)^{-2} tr(\mathbf{K}^x\mathbf{AK}^y\mathbf{A})\\
    s.t. \  \mathbf{K}_{ab}^{v} = \kappa(\mathbf{P}_{a \cdot}^{v},\mathbf{P}_{b \cdot}^{v}) = e^{(\frac{-\begin{Vmatrix} \mathbf{P}_{a \cdot}^{v} - \mathbf{P}_{b \cdot}^{v} \end{Vmatrix}^2}{\sigma^{v}})}
     \end{aligned}
\end{equation}
where $\mathbf{K}^x$ and $\mathbf{K}^y$ are the kernel-induced similarity matrix from $\mathbf{P}^x$ and $\mathbf{P}^y$, respectively. $\mathbf{A}$ is a centering matrix: $\mathbf{A} = \mathbf{I}-\mathbf{ee}^{\mathrm{T}}/n$, where $\mathbf{e} = (1,\cdots,1)^{\mathrm{T}} \in \mathbb{R}^n$ and $\mathbf{I}$ is the identity matrix. 

For the decoding part, we use the extracted commonality $\mathbf{C}^*$ shared across modalities and the individuality of each modality to reconstruct this original modality as follows:
\begin{equation}
\mathcal{J}_3(\mathbf{C}^*, \mathbf{P}^{v}) = {\sum}_{v=x,y}\frac{\begin{Vmatrix} \mathbf{F}^{v}-f_{de}^v([\mathbf{C}^*,\mathbf{P}^{v}])  \end{Vmatrix}_F^2}{n d_{v}}
\end{equation}
where $\{f_{de}^{v}\}$ is the corresponding decoder of each modality.

Then we can define the loss function of the individuality-commonality AE as:
\begin{equation}
\label{eq9}
\mathop{\min}_{\theta_z} Loss1 = \alpha \mathcal{J}_1(\mathbf{C}^*) + \beta \mathcal{J}_2(\mathbf{P}^{v})+\mathcal{J}_3(\mathbf{C}^*, \mathbf{P}^{v})
\end{equation}
$\theta_z$ are the parameters of the individuality-commonality AE.  $\alpha$ and $\beta$ ($\in (0,1] $) are the parameters to control the weight of individuality and commonality information. To this end, LtCMH can capture the commonality and individuality of different modalities, which will be used to create dynamic meta features in the next subsection.


\subsection{Dynamic Meta Features Learning}
For long-tail datasets, the head labels have abundant representations while the tail labels don't \cite{cui2019class}. As a result, head labels can be easily distinguished from each other in the feature space, but tail labels can not. To enrich label representation and transfer knowledge of head labels to tail ones, we propose the dynamic meta memory embedding module using direct features of $\mathbf{F}^x$ and $\mathbf{F}^y$, and the commonality $\mathbf{C}^*$ and individuality $\mathbf{P}^{v}$. 

The modality heterogeneity is a typical issue in cross modal information fusion, which means that different modalities may have completely different feature representations and statistical characteristics for the same semantic labels. This makes it difficult to directly measure the similarity between different modalities. However, multi-modal data that describe the same instance usually have close semantic meanings. In the previous subsection, we have learnt a commonality information matrix $\mathbf{C}^*$ by mining the geometric relations between samples of two labels in the embedding space. Therefore, $\mathbf{C}^*$ can be seen as a bridge between different modalities.
Meanwhile, different modalities often have their own individual information, so we learn $\mathbf{P}^{v}$ to capture the individuality of each modality. We take $\mathbf{C}^*$ and $\mathbf{P}^{v}$ as the memory features, and obtain the meta features of a modality as follows:
\begin{equation}
\label{eq8}
    \mathbf{M}^{v} = \mathbf{F}^v + \mathbf{E}_1\odot \mathbf{C^*} + \mathbf{E}_2 \odot \mathbf{P}^{v}
\end{equation}
Data-poor tail labels need more memory features to enrich their representations than data-rich head labels, we design two adaptive selection factors $\mathbf{E}_1$ and $\mathbf{E}_2$ on $\mathbf{C^*}$ and $\mathbf{P}^{v}$ for this purpose, {which can be adaptive computed from $\mathbf{F}^v$ to derive selection weights as $\mathbf{E}_1=Tanh(FC(\mathbf{F}^v))$ and $\mathbf{E}_2=Tanh(FC(\mathbf{F}^v))$. $\odot$ is the Hadamard product. To match the general multi-modal data in a soft manner, we adopt the lightweight `Tanh+FC' to avoid complex and inefficient parameters adjusting.}
Different from long-tail hashing methods \cite{kouattention} that use prototype networks \cite{snell2017prototypical} or simply stack the direct features to learn meta features for each label, our $\mathbf{M}^v$ is created from both individuality and commonality, and direct features of multi-modal data. In addition, $\mathbf{M}^v$ is more interpretable and effective, and does not need the clear division of head and tail labels.

\subsection{Hash Code Learning}
After obtaining the dynamic meta features, the information across modalities is reserved and each sample's representation is enriched. Alike DCMH \cite{jiang2017deep}, we use $\mathbf{S}\in \mathbb{R}^{n\times n}$ to store the similarity of $n$ training samples. $S_{ij}=1$ means that $\mathbf{x}_i$ and $\mathbf{y}_j$ are with the same label, and $S_{ij}=0$ otherwise. Based on $\mathbf{M}^x$, $\mathbf{M}^y$ and $\mathbf{S}$, we can define the likelihood function as follows:
\begin{equation}
    p(\mathbf{S}|\mathbf{M}^{x},\mathbf{M}^{y}) = \begin{cases}    \sigma(\mathbf{\phi}^{xy}) &  S_{ij}=1 \\ 1- \sigma(\mathbf{\phi}^{xy}) &  S_{ij}=0          \end{cases}
\end{equation}
where $\mathbf{\phi}^{xy}=1/2(\mathbf{M}^{x})^{\mathrm{T}}\mathbf{M}^{y}$ and {$\sigma(\mathbf{\phi}^{xy}) = \frac{1}{1+e^{-\mathbf{\phi}^{xy}}}$. The smaller the angle between $\mathbf{M}^x$ and $\mathbf{M}^y$ is, the larger the $\phi^{xy}$ is, which makes it a higher probability that $S_{ij}= 1$ and vice versa.} Then we can define the loss function for learning hash codes as follows:
\begin{equation}
    \begin{aligned}
    \min_{\mathbf{B},\theta_x,\theta_y} Loss2&=-\sum_{i,j=1}^n(S_{ij}\phi_{ij}^{xy}-\log(1+e^{\phi_{ij}^{xy}}))\\
   +\gamma (\begin{Vmatrix} \mathbf{B}-\mathbf{M}^{x} \end{Vmatrix}_F^2 &+ \begin{Vmatrix} \mathbf{B}-\mathbf{M}^{y}  \end{Vmatrix}_F^2) +\eta (\begin{Vmatrix} \mathbf{M}^{x}\mathbf{1}  \end{Vmatrix}_F^2 + \begin{Vmatrix} \mathbf{M}^{y}\mathbf{1}  \end{Vmatrix}_F^2) \\
    & s.t.\quad \mathbf{B}\in\{+1,-1\}^{c\times n}&
    \end{aligned}
    \label{eq14}
\end{equation}
$\phi_{ij}^{xy}$ means the $i$-th row and $j$-th column of $\mathbf{\phi}^{xy}$. $\mathbf{B}$ is the unified hash codes and $\mathbf{1}$ is a vector with all elements being 1. $\theta_x$ and $\theta_y$ are the parameters of image and text meta feature learning module. The first term is the negative log likelihood function, it aims at minimizing the inner product between $\mathbf{M}^{x}$ and $\mathbf{M}^{y}$ and preserving the semantic similarities of samples from different modalities. The Hamming distance between similar samples will be reduced and between dissimilar ones will be enlarged. The second term aims to minimize the difference between the unified hash codes and meta features of each modality. Due to the similarity preservation of $\mathbf{M}^x$ and $\mathbf{M}^y$ in $\mathbf{S}$, enforcing the unified hash codes closer to $\mathbf{M}^x$ and $\mathbf{M}^y$ is expected to preserve the cross modal similarity to match the goal of cross modal hashing. The last term pursues balanced binary codes  with fixed length for a larger coding capacity.

\subsection{Optimization}
There are three parameters $\theta_x$, $\theta_y$, $\mathbf{B}$ in our hash learning loss function in Eq. \eqref{eq14}, it is difficult to simultaneously optimize them all and find the global optimum. Here, we adopt a canonically-used alternative strategy that optimizes one of them with others fixed, and give the optimization as below.

\textbf{Optimize $\theta_x$ with fixed $\theta_y$ and $\mathbf{B}$}: We first calculate the derivative of the loss function ${Loss2}$ with respect to $\mathbf{M}^{x}$ for image modality and then we take the back-propagation (BP) and stochastic gradient decent (SGD) to update $\theta_x$ until convergence or the preset maximum epoch.
$\frac{\partial {Loss2}}{\partial \mathbf{M}^{x}}$ can be calculated as:
\begin{equation}
    \begin{aligned}
    \frac{\partial {Loss2}}{\partial \mathbf{M}^{x}_{*i}} =& \frac{1}{2} {\sum}_{j=1}^{n}(\sigma(\phi_{ij})\mathbf{M}_{*j}^y-S_{ij}\mathbf{M}^y_{*j})\\
    & +2\gamma(\mathbf{M}^{x}_{*i}-\mathbf{B}_{*i})+2\eta \mathbf{M}^{x}\mathbf{1}
    \end{aligned}
    \label{eq11}
\end{equation}
where $\mathbf{M}^{x}_{*i}$ represents the $i$-th column of $\mathbf{M}^{x}$. By the chain rule, we update $\theta_x$ based on BP algorithm.

The way to optimize $\theta_y$ is similar as that to optimize $\theta_x$.

\textbf{Optimize $\mathbf{B}$ with fixed $\theta_x$ and  $\theta_y$}:
When $\theta_x$ and  $\theta_y$ are fixed, the optimization problem can be reformulated as:
\begin{equation}
\begin{aligned}
\max_\mathbf{B} &\quad tr(\mathbf{B}(\mathbf{M}^{x}+\mathbf{M}^{y})^T)\\
    s.t. &\quad \mathbf{B}\in \{+1,-1\}^{c\times n}
\end{aligned}
\end{equation}
We can derive that the optimized $\mathbf{B}$ has the same sign as $(\mathbf{M}^{x}+\mathbf{M}^{y})$:
\begin{equation}
\label{eq18}
   \mathbf{B}= sign(\mathbf{M}^{x}+\mathbf{M}^{y})
\end{equation}

We illustrate the whole framework of LtCMH in Figure \ref{fig:framework}, and defer its algorithmic procedure in the Supplementary file.

\section{Experiments}
\subsection{Experimental setup}
There is no off-the-shelf benchmark long-tail multi-modal dataset for experiments, so we pre-process two hand-crafted multi-modal datasets (Flickr25K \cite{huiskes2008mir} and NUS-WIDE \cite{chua2009nus}), to make them fit long-tail settings. The statistics of pre-processed datasets are reported in Table \ref{table:dataset}, and more information of the pre-processings are given in the Supplementary file. We also take the public Flickr25K and NUS-WIDE as the balanced datasets for experiments.

\begin{table}[h!]
\small
    \begin{center}
        \caption{Statistics of long-tail datasets. $z_a$ is the number of samples annotated with label $a$, which conforms to Zipf's law distribution; $c$ is the number of distinct labels.}
        \label{table:dataset}
        \label{table:sample}
        \begin{tabular}{c| r r r r}
             \hline
             Dataset & $N_{base}$ & $N_{query}$  & $z_1$ ($z_c$) &c\\
             \hline
             Flicker25K & 18015 &2000  & 3000(60) &24\\
            \hline
            NUS-WIDE & 195834 &2000  & 5000(100) &21\\
            \hline
        \end{tabular}
    
    \end{center}
\end{table}


Alike DCMH \cite{jiang2017deep}, we use a pre-trained CNN named CNN-F with 8 layers, to extract the direct image features $\mathbf{F}^x$, and another network with two fully connected layers to extract direct text features $\mathbf{F}^y$. These two networks are with the same hyper-parameters as DCMH. Other networks can also be used here, which is not the main focus of this work.
SGD is used to optimize model parameters. Learning rate of image and text feature extraction is set as 1e-1.5, the learning rate of individuality-commonality AE is set as 1e-2. Other hyper-parameters are set as: batch size=128, $\alpha$=0.05, $\beta$=0.05, $\gamma$=1, $\eta$=1, $d_x$ and $d_y$ is equal to hash code length $k$, the max epoch is 500. Parameter sensitivity is studied in the Supplementary file.

We compare LtCMH against with six representative and related CMH methods, which include CMFH \cite{ding2014collective}, JIMFH \cite{wang2020joint}, DCMH \cite{jiang2017deep}, DGCPN \cite{yu2021deep}, SSAH \cite{li2018self}, and MetaCMH \cite{wang2021meta}. The first two are shallow solutions, and latter four are deep ones. They all focus on the commonality of multi-modal data to learn hash codes.  CMFH, JIMFH and DGCPN are unsupervised solutions, while the others are supervised ones.
We also take the recent long-tail single-modal hashing method LTHNet \cite{chen2021long} as another baseline. For fair evaluation with LTHNet, we solely train and test LtCMH and LTHNet on the image-modality. The parameters of compared method are fixed as reported in original papers or selected by a validation process. All experiments are independently repeated for ten times. Our codes will be made public later.

\subsection{Result Analysis}
\begin{table*}[t]

\centering
\caption{Results (MAP) of each method on \emph{long-tail} Flickr25K and NUS-WIDE. The best results are in \textbf{boldface}.}
\small
\label{table1}
\begin{tabular}{c|c|ccc|ccc}
\hline
& &\multicolumn{3}{c|}{Flicker25K}&\multicolumn{3}{c}{NUS-WIDE}\\
\hline
 & & 16-bits & 32-bits & 64-bits & 16-bits & 32-bits & 64bits\\
\hline
\multirow{6}{*}{I$\rightarrow$T}  
& CMFH & .354$\pm$.030  & .377$\pm$.005 &  .382$\pm$.003 & .254$\pm$.010 & .256$\pm$.007 & .261$\pm$.020\\
& JIMFH& .400$\pm$.009 & .414$\pm$.010 & .433$\pm$.026& .308$\pm$.032 & .323$\pm$.027 &.347$\pm$.016 \\
& DGCPN& .538$\pm$.008  & .551$\pm$.013 &.579$\pm$.005  & .379$\pm$.006 & .381$\pm$.007 & .409$\pm$.004\\
& DCMH& .477$\pm$.020 & .492$\pm$.003 & .506$\pm$.017  & .347$\pm$.014 &.367$\pm$.003  & .376$\pm$.010 \\
& SSAH& .571$\pm$.004  & .588$\pm$.005 & .603$\pm$.009 & .371$\pm$.007 & .383$\pm$.006 & .418$\pm$.004\\
& MetaCMH& .608$\pm$.016  & .621$\pm$.009& .624$\pm$.025 &.409$\pm$.016  &.421$\pm$.004  &.430$\pm$.007 \\
\cline{2-8}
& LtCMH &\bf .687$\pm$ .015  & \bf.732$\pm$.004 & \bf.718$\pm$.014 & \bf.433$\pm$.004  & \bf.475$\pm$.008  & \bf.532$\pm$.017\\
\hline
\multirow{6}{*}{T$\rightarrow$I}  
& CMFH & .366$\pm$.015  & .382$\pm$.005 & .396$\pm$.010 & .269 $\pm$.025 & .279$\pm$.004 & .287$\pm$.001\\
& JIMFH& .433$\pm$.008& .449$\pm$.007 & .448$\pm$.014& .368$\pm$.009 &.372$\pm$.018 & .379$\pm$.020\\
& DGCPN& .529$\pm$.014& .541$\pm$.007 & .577$\pm$.016 & .377$\pm$.003  & .388$\pm$.008 &.420$\pm$.012 \\
& DCMH& .500$\pm$.010  & .510$\pm$.007 & .514$\pm$.005 & .348$\pm$.020& .380$\pm$.008 & .401$\pm$.009\\
& SSAH& .566$\pm$.012 & .579$\pm$.006 & .630$\pm$.008 &.382$\pm$.005  & .415$\pm$.013 &.426$\pm$.007 \\
& MetaCMH& .624$\pm$.002 & .640$\pm$.006 & .643$\pm$.032 & .416$\pm$.004 & .433$\pm$.003 &.438$\pm$.009 \\
\cline{2-8}
& LtCMH & \bf.729$\pm$.008  & \bf.738$\pm$.015 & \bf.750$\pm$.006 & \bf.441$\pm$.008  & \bf.458$\pm$.012 & \bf.463$\pm$.006\\
\hline

\end{tabular}
\end{table*}

\begin{table*}[t]
\centering
\small
\caption{Results (MAP) of each method on \emph{ balanced} Flickr25K and NUS-WIDE. The best results are in \textbf{boldface}.}
\label{table2}
\begin{tabular}{c|c|ccc|ccc}
\hline
& &\multicolumn{3}{c|}{Flicker25K}&\multicolumn{3}{c}{NUS-WIDE}\\
\hline
 & & 16-bits & 32-bits & 64-bits & 16-bits & 32-bits & 64bits\\
\hline
\multirow{6}{*}{I$\rightarrow$T}  & CMFH &  .597$\pm$.037 & .597$\pm$.036 & .597$\pm$.037 & .409$\pm$.041 & .419$\pm$.051 & .417$\pm$.062\\
& JIMFH& .621$\pm$.038 & .635$\pm$.050 &.635$\pm$.045 & .503$\pm$.051 & .524$\pm$.062 & .529$\pm$.064\\
& DGCPN& .732$\pm$.010  &.742$\pm$.004  & .751$\pm$.008 &.625$\pm$.003  & .635$\pm$.007 &.654$\pm$.020 \\
& DCMH&  .710$\pm$.022 & .721$\pm$.017 & .735$\pm$.014 & .573$\pm$.027 & .603$\pm$.003 & .609$\pm$.002\\
& SSAH& .738$\pm$.018  & .750$\pm$.010 & .779$\pm$.009 & .630$\pm$.001 & .636$\pm$.010 & .659$\pm$.004\\
& MetaCMH & .708$\pm$.007  &.716$\pm$.005 & .724$\pm$.013 & .612$\pm$.011 & .619$\pm$.005 & .644$\pm$.018\\
\cline{2-8}
& LtCMH & \bf.745$\pm$.018 & \bf.753$\pm$.012 & \bf.781$\pm$.011 & \bf.635$\pm$.010  & \bf.654$\pm$.020 & \bf.678$\pm$.003\\
\hline
\multirow{6}{*}{T$\rightarrow$I}  & CMFH &  .598$\pm$.048 & .602$\pm$.055 & .601$\pm$.065 & .412$\pm$.072 & .417$\pm$.061 & .418$\pm$.068\\
& JIMFH&  .650$\pm$.042 & .662$\pm$.064 &.657$\pm$.059 & .584$\pm$.081 & .604$\pm$.052& .655$\pm$.069\\
& DGCPN& .729$\pm$.015 &.741$\pm$.008  &.749$\pm$.014  & .631$\pm$.008 & .648$\pm$.013 & .660$\pm$.004\\
& DCMH&  .738$\pm$.020 & .752$\pm$.020 &.760$\pm$.019 & .638$\pm$.010 & .641$\pm$.011 & .652$\pm$.012\\
& SSAH& .750$\pm$.028 & .795$\pm$.023 & .799$\pm$.008 & \bf.655$\pm$.002 & \bf.662$\pm$.012 & .669$\pm$.009\\
& MetaCMH& .741$\pm$.009 & .758$\pm$.014 & .763$\pm$.004 & .594$\pm$.004& .611$\pm$.007 & .649$\pm$.008\\
\cline{2-8}
& LtCMH & \bf.770$\pm$.008  & \bf.795$\pm$.004 & \bf.802$\pm$.023 & .608$\pm$.029  & .644$\pm$.010 & \bf.688$\pm$.007\\
\hline
\end{tabular}
\label{table_MAP}
\end{table*}

\begin{figure}[hbtp]
    \centering
    \includegraphics[width=8cm]{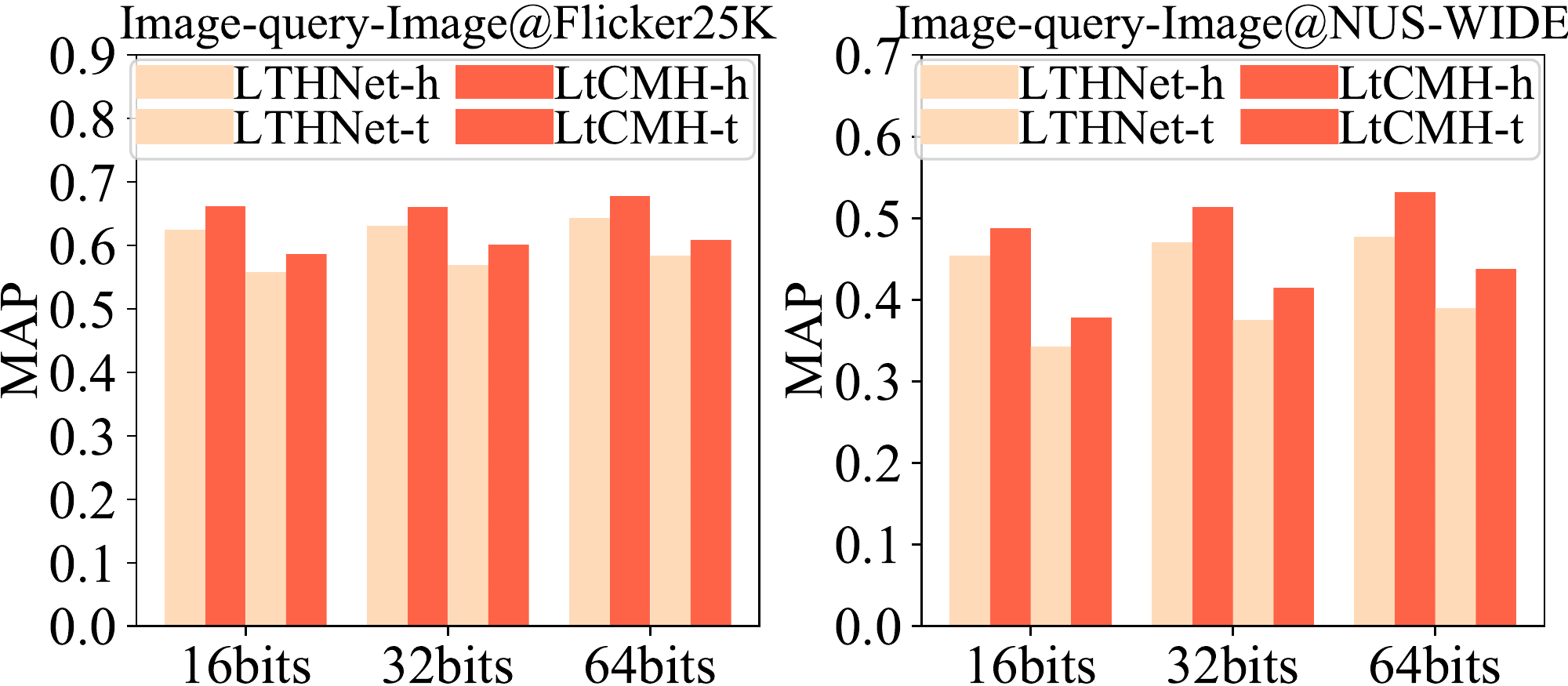}
    \caption{Performance comparison of LtCMH and LTHNet on \emph{head} and \emph{tail} labels.  For Flicker, the first 14 labels are head; and for NUS-WIDE, the first 15 labels are head.}
    \label{fig:htL}
\end{figure}

We adopt the typical mean average precision (MAP) as the evaluation metric, and report the average results and standard deviation in Table \ref{table1} (long-tailed), Table \ref{table2} (balanced) and Table \ref{table3} (single-modality). From these results, we have several important observations:\\
{
(i) LtCMH can effectively handle long-tail multi/single-modal data, this is supported by the clear performance gap between LtCMH and other compared methods. DCMH and SSAH have greatly compromised  performance on long-tailed datasets, because they use semantic labels to guide hash codes learning, while tail labels do not have enough training samples to preserve the modality relationships in both the common semantic space and Hamming space. 
Although unsupervised CMH methods give hash codes without referring to the skewed label distributions, they are also misled by the overwhelmed samples of head labels. Another cause is that they all focus the commonality of multi-modal data, while LtCMH considers both the commonality and individuality. 
We note each compared method has a greatly improved performance on the balanced datasets, since they all target at balanced multi-modal data. Compared with results on long-tail datasets, the performance drop of LtCMH is the smallest, this proves its generality. 
LtCMH sometimes slightly loses to SSAH on the balanced datasets, that is because the adversarial network of SSAH can utilize more semantic information in some cases.\\
(ii) LtCMH can learn more effective meta features and achieve better knowledge transfer from head labels to tail labels than LTHNet and MetaCMH. The latter two stack direct features as memory features to transfer knowledge from head to tail labels, they perform better than other compared methods (except LtCMH), but the stacked features may cause information overlap and they only enrich tail labels within each modality, and suffer the inconsistency caused by modality heterogeneity. LtCMH captures both individuality and commonality of multi-modal data, and the enhanced commonality helps to keep cross-modal consistency. 
{We further separately measure the performance of LtCMH, MetaCMH and LTHNet on the head labels and tail ones, and report the results in Figure \ref{fig:htL} and Supplementary file. We find that LtCMH gives better results than MetaCMH and LTHNet on both head and tail labels. These results further suggest the effectiveness of our  meta features.}  \\
(iii) The consideration of label information and data distribution improve the performance of CMH. We find that most supervised methods perform better than unsupervised ones.
As an exception, DGCPN builds three graphs to comprehensively explore the structure information of multi-modal data, and thus gives a better performance than the supervised DCMH. Deep neural networks based solutions also perform better than shallow ones. Besides deep feature learning, SSAH and MetaCMH leverage both the label information and data distribution, they obtain a better performance than other compared methods. Compared with SSAH and MetaCMH, LtCMH leverages these information sources in a more sensible way, and thus achieves a better performance on long-tail and balanced datasets.
}

\begin{figure*}[h!tbp]
\subfigure[Flicker25K]{
\includegraphics[width=9cm]{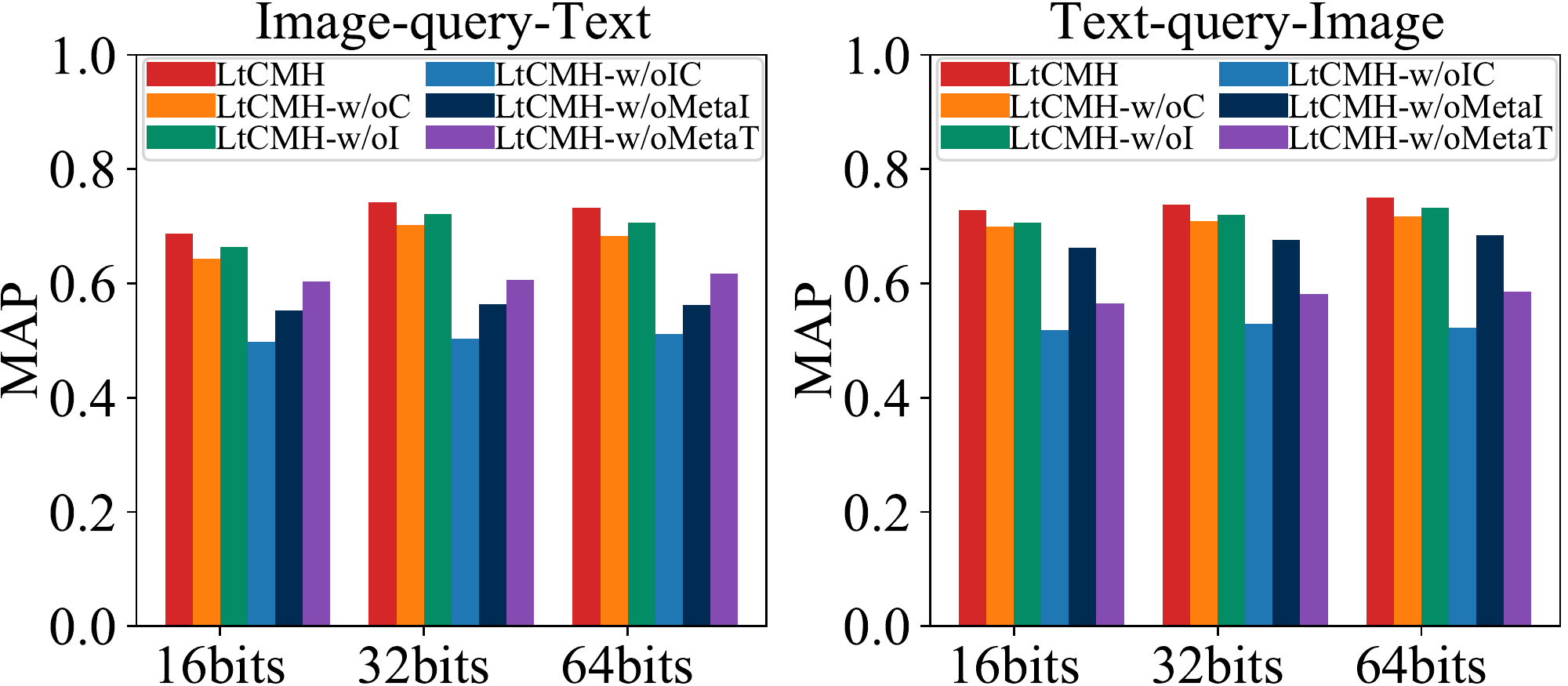}}
\subfigure[NUS-WIDE]{
\includegraphics[width=9cm]{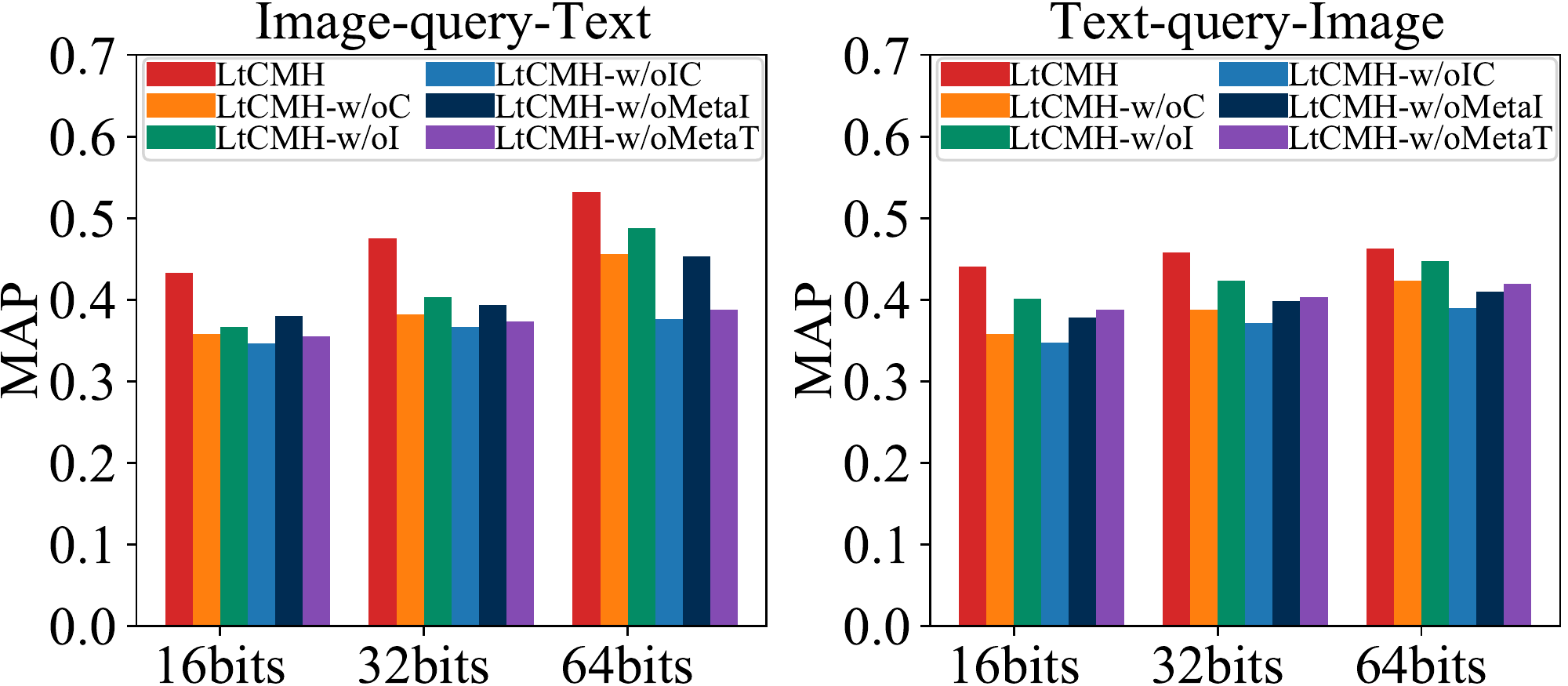}}
\caption{Results of LtCMH and its variants on \emph{long-tailed} Flicker25K and NUS-WIDE.}
\label{fig:ablation}
\end{figure*}

\begin{table}[hbtp]
\small
    \centering
    \caption{Results of LtCMH and LTHNet on  \emph{long-tailed} Flickr25K and NUS-WIDE. Better results are in \textbf{boldface}. }
    \label{table3}
    \begin{tabular}{|c|c|c|c|c|}
    \hline
         & & 16bits & 32bits &64bits  \\
    \hline
         \multirow{2}{*}{Flickr25K}& LTHNet & .602 & .573 & .610 \\
         & LtCMH &\bf.651 &\bf.647 & \bf.654 \\
         \hline
        \multirow{2}{*}{NUS-WIDE}& LTHNet &.401 & .411& .420\\
         & LtCMH & \bf.473& \bf.508 & \bf.513 \\
         \hline
    \end{tabular}
    \label{tab:my_label}
\end{table}

\subsection{Ablation Experiments}
{To gain an in-depth understanding of LtCMH, we introduce five variants: LtCMH-w/oC, LtCMH-w/oI, LtCMH-w/oIC, LtCMH-w/oMetaI, LtCMH-w/oMetaT, which separately disregards the individuality, commonality and the both, dynamic meta features from the image modality and text modality. Figure \ref{fig:ablation} shows the results of these variants on Flicker25K and NUS-WIDE. We have several important observations. (i) Both the commonality and individuality are important for LtCMH on long-tail datasets. This is confirmed by clearly reduced performance of LtCMH-w/oC and LtCMH-w/oI, and by the lowest performance of LtCMH-w/oIC. We find LtCMH-w/oI performs better than LtCMH-w/oC, since commonality captures the shared and complementary information of multi-modal data, and bridges two modalities for cross-modal retrieval. 
(ii) Both the meta features from the text- and image-modality are helpful for hash code learning. We note a significant performance drop when meta features of target query modality are unused.


Besides, we also study the impact of input parameters $\alpha$, $\beta$, $\gamma$ and $\eta$, which control the learning loss of commonality, individuality, consistency of hash codes across modalities, balance of hash codes. We find that a too small $\alpha$ or $\beta$ cannot ensure to learn commonality and individuality across modalities well, but a too large of them brings down the validity of AE. A too small $\gamma$ and $\eta$ cannot keep consistency of hash codes across modalities and generate balanced hash codes. Given these results, we set $\alpha$=0.05, $\beta$=0.05, $\gamma$=1, $\eta$=1.
}

\section{Conclusion}
In this paper, we study how to achieve CMH on the prevalence long-tail multi-modal data, which is a practical and important, but largely unexplored problem in CMH. We propose an effective approach LtCMH that leverages the individuality and commonality of multi-modal data to create dynamic meta features, which enrich the representations of tail labels and give discriminant hash codes. The effectiveness and adaptivity of LtCMH are verified by experiments on long-tail and balanced multi-modal datasets.

\newpage

\section{Long-tail Cross Modal Hashing \\
Supplementary file}

\subsection{Algorithm Table}
Algorithm \ref{alg:Algorithm} summarizes the whole learning procedure of LtCMH. Lines 3-8 train the individuality-commonality AE on multi-modal data; lines 10-15 learn dynamic meta features using individuality, commonality and direct features of respective modalities; and line 16 quantifies the meta features to generate hash codes.

\begin{algorithm}[ht]
\caption{\textbf{LtCMH}: Long-tail Cross-Modal Hashing}
\label{alg:Algorithm}
\begin{algorithmic}[1] 
\REQUIRE Image set $\mathcal X$, Text set $\mathcal Y$, Similarity matrix $\mathbf{S}$, the maximum number of epochs $maxE$ and the hyperparameters $\alpha$, $\beta$, $\gamma$ and $\eta$.
\ENSURE $\theta_x$, $\theta_y$ for hashing networks, $\theta_z$ for auto-encoders and $\mathbf{B}$ for hash codes.\\
\STATE Initialize $\theta_x$, $\theta_y$, load $\theta_z$ of autoencoder,  batch size $N=128$, iteration number $t = \lceil n/N\rceil$.\\
\STATE $epoch=0$;
    \WHILE{ $epoch++ < maxE$ }
      \FOR{$iter = 1,2,\cdots,t$}
            \STATE Randomly sample $N$ points from $\mathcal X$ and $\mathcal Y$, then calculate $\mathbf{P}^x, \mathbf{P}^y$ according to Eq. (1), $\mathbf{C^*}$ according to Eq. (2) in the mini-batch by forward propagation.\
            \STATE $\theta_z$ = SGD(Loss1,$\theta_z$).
      \ENDFOR
    \ENDWHILE
    \STATE $epoch=0$;
    \WHILE{$epoch++ < maxE$ }
      \FOR{$iter = 1,2,\cdots,t$}
            \STATE Randomly sample $N$ points from $\mathcal{X}$/$\mathcal{Y}$, then calculate $\mathbf{M}^x$/$\mathbf{M}^y$ according to Eq. (8) in the mini-batch by forward propagation.\
            \STATE Calculate the derivative according to Eq. (11).
            \STATE Update the parameter $\theta_x$/$\theta_y$ by  back propagation.
      \ENDFOR
      \STATE Learn $\mathbf{B}$ by Eq. (13).
    \ENDWHILE
\end{algorithmic}
\end{algorithm}

\subsection{Configuration of Image and Text Networks}
Alike DCMH \cite{jiang2017deep}, we adopt a pre-trained CNN with eight layers called CNN-F to extract image feature. The first five layers are convolutional layers denoted by `convolution1-5'. `filters: $num$ $\times$ $size$ $\times$ $size$' means the layer has $num$ convolution  filters and each filter size is $size \times size$. `st.t' means the convolution stride is $t$. `pad n' means adding $n$ pixels to each edge of the input image. `LRN' means Local Response Normalization. `pool' is the down-sampling factor. `fc' means the fully connect layer. `nodes:n' means there are $n$ nodes in the layer. $k$ is the length of hash codes. For text modality, we adopt a neural network with two fully connected layers (FC). The detailed structures of CNN-F and FC are shown in Table \ref{implementation detail}.  
\begin{table}[h!]

    \begin{center}
        \caption{Configuration of Image and Text Networks}
        \label{implementation detail}
        \begin{tabular}{c|c|c}
             \hline
             Network & Layers & Configuration \\
             \hline
             \multirow{13}{*}{Image-Extract} & \multirow{2}{*}{convolution1} & filters:64 × 11 × 11, st. 4, \\ &&pad 0,LRN, × 2 pool\\
             \cline{2-3}
             & \multirow{2}{*}{convolution2} & filters:256 × 5 × 5, st. 1, \\ &&pad 2,LRN, × 2 pool\\
             \cline{2-3}                                                                        
             & \multirow{2}{*}{convolution3} & filters:256 × 3 × 3,\\ &&st. 1, pad 1 \\
             \cline{2-3}
             & \multirow{2}{*}{convolution4} & filters:256 × 3 × 3,\\ &&st. 1, pad 1 \\
             \cline{2-3}
             & \multirow{2}{*}{convolution5} & filters:256 × 3 × 3, st. 1,\\ &&  pad 1,×2 pool\\
             \cline{2-3} 
             & \multirow{1}{*}{fc6} & nodes:4096\\
             \cline{2-3}
             & \multirow{1}{*}{fc7} & nodes:4096\\
             \cline{2-3}
             & \multirow{1}{*}{fc8} & nodes:k\\
             \cline{1-3}
             \multirow{2}{*}{Text-Extract} &fc1 & nodes:8192\\
             \cline{2-3}
             & fc2 &nodes:k\\
             \cline{1-3}
             
        \end{tabular}

    \end{center}
\end{table}

\subsection{Dataset preprocessing}

There are 25000 images in Flickr25K, and each image has its corresponding text of 24 different semantic labels. In our experiments, we found some of the texts or labels in Flicker25K are all zero vectors. To be fair, we follow the previous works that only retained the samples having at least 20 rows of text and we finally get a cleaned dataset with 20015 samples. Each image's corresponding textual descriptions are processed into BOW vectors with 1386 dimensions. We use each image-text pair as a sample for training. We randomly sampled 2000 samples as the query(test) set, and the remaining 18015 samples as database(retrieval) set.

NUS-WIDE dataset contains 260648 image-text pairs, each pair of which is tagged with 21 different semantic labels. We follow the settings of DCMH \cite{jiang2017deep} to select 195834 image-text pairs which belong to 21 most frequent concepts. We then processed the texts into 5018-dimentional BOW vectors. We randomly sampled 2000 samples as the query(test) set, and the remaining samples as database(retrieval) set. In addition, the ground-truth neighbors are defined as those image-text pairs which share at least one common label for Flicker25K or NUS-WIDE.

The simple preprocessing  described above is not enough to obtain a usable long-tailed dataset. Because most samples in both datasets have more than one label. For those samples with more than 3 labels, we select 2 labels with the most discrimination. What we called a discriminant label is the label with fewer samples. Then we follow Zipf’s law to randomly sample the training set from database, building the long-tailed dataset according to imbalance factor as 50. 

    

\subsection{Additional Result and Analysis}
\begin{figure}[h!tbp]
    \centering
    \includegraphics[width=8cm]{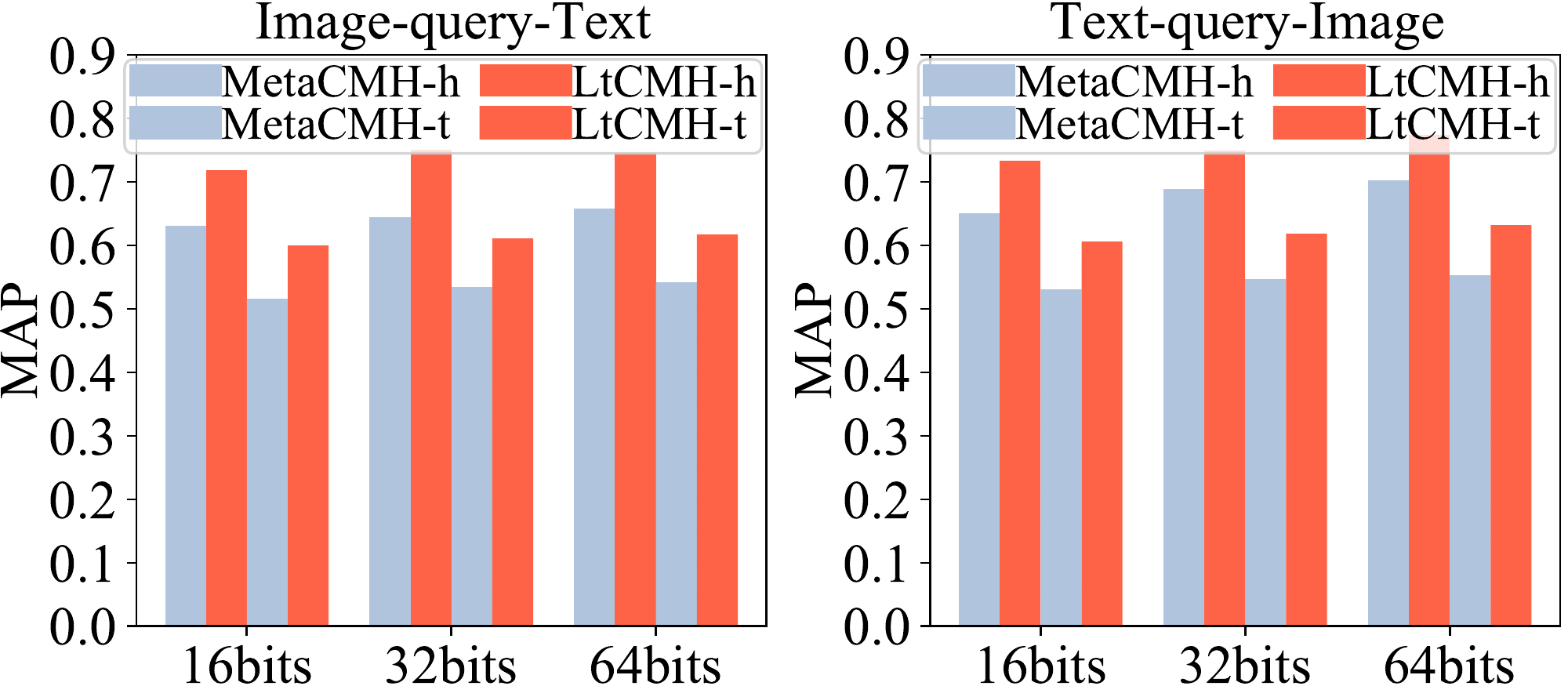}
    \caption{Performance comparison of LtCMH and MetaCMH on \emph{head} and \emph{tail} labels on Flicker25K. The first 14 labels are head.}
    \label{fig:htMf}
\end{figure}
\begin{figure}[h!tbp]
    \centering
    \includegraphics[width=8cm]{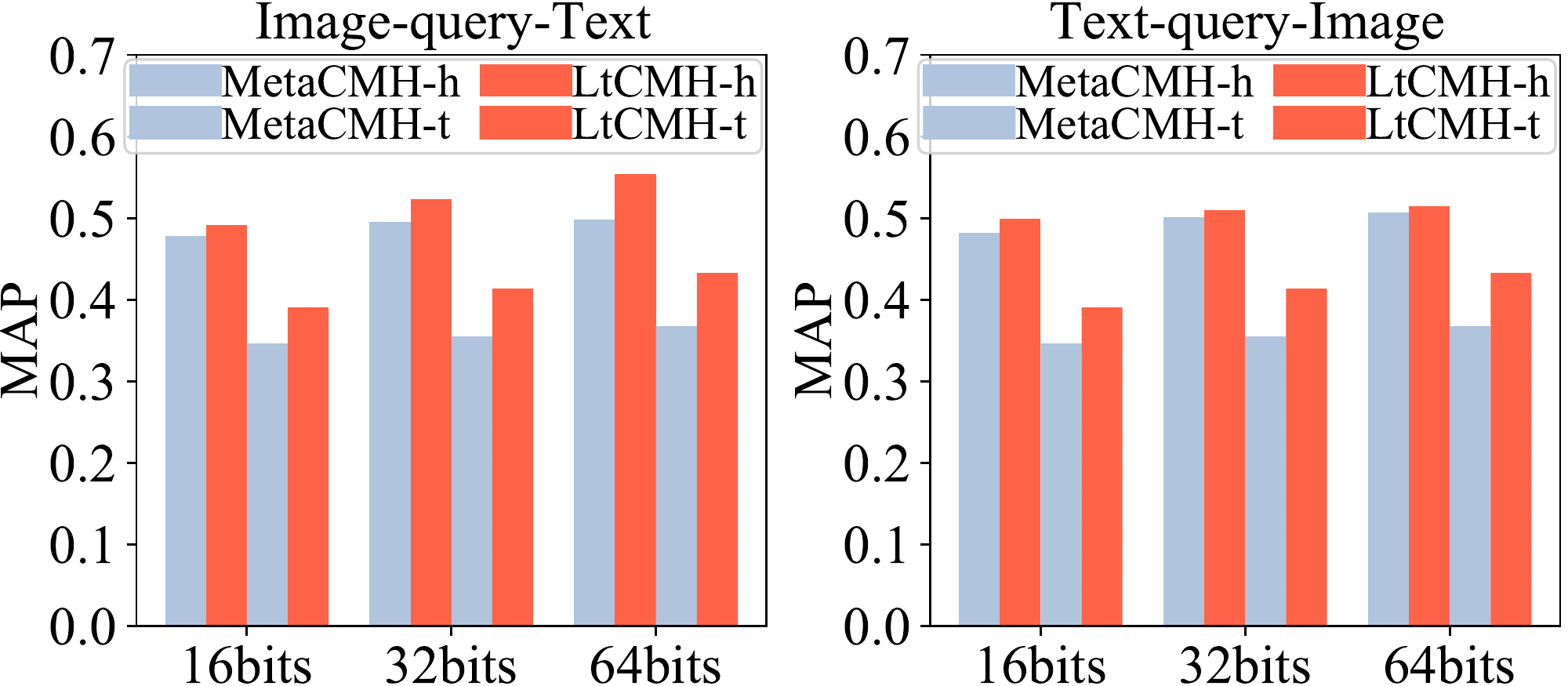}
    \caption{Performance comparison of LtCMH and MetaCMH on \emph{head} and \emph{tail} labels on NUS-WIDE. The first 15 labels are head.}
    \label{fig:htMn}
\end{figure}
Figure \ref{fig:htMf} and Figure \ref{fig:htMn} report the results of MetaCMH and LtCMH on the head and tail labels on Flicker25K and NUS-WIDE. Here, LtCMH-h (MetaCMH-h) means LtCMH (MetaCMH) on head labels, and LtCMH-t (MetaCMH-t) denotes LtCMH (MetaCMH) on tail labels.

We find that LtCMH gives better results than MetaCMH on both head and tail labels on both datasets.
MetaCMH simply stacks direct features which may cause information overlap and only enriches tail labels within each modality, and suffers the inconsistency caused by modality heterogeneity. LtCMH captures both individuality and commonality of multi-modal data, thus utilizing the commonality to pursue cross-modal consistency. Both the individuality and commonality model the complex relation between labels and multi-modal data. For these advantages, it outperforms MetaCMH on both head labels and tail ones. This proves the generalization of our dynamic meta features.

\subsection{Parameter Sensitivity Analysis}
There are four input parameters involved with LtCMH including $\alpha$, $\beta$, $\gamma$ and $\eta$, which separately control the learning loss of commonality, individuality, consistency of hash codes across modalities, and balance of hash codes. Following the previous experimental protocol, we conduct experiments to study the impact of each parameter by varying its values while fixing the other parameters.  

\begin{figure}[h!tbp]
\subfigure[Flicker25K]{
\includegraphics[width=8cm]{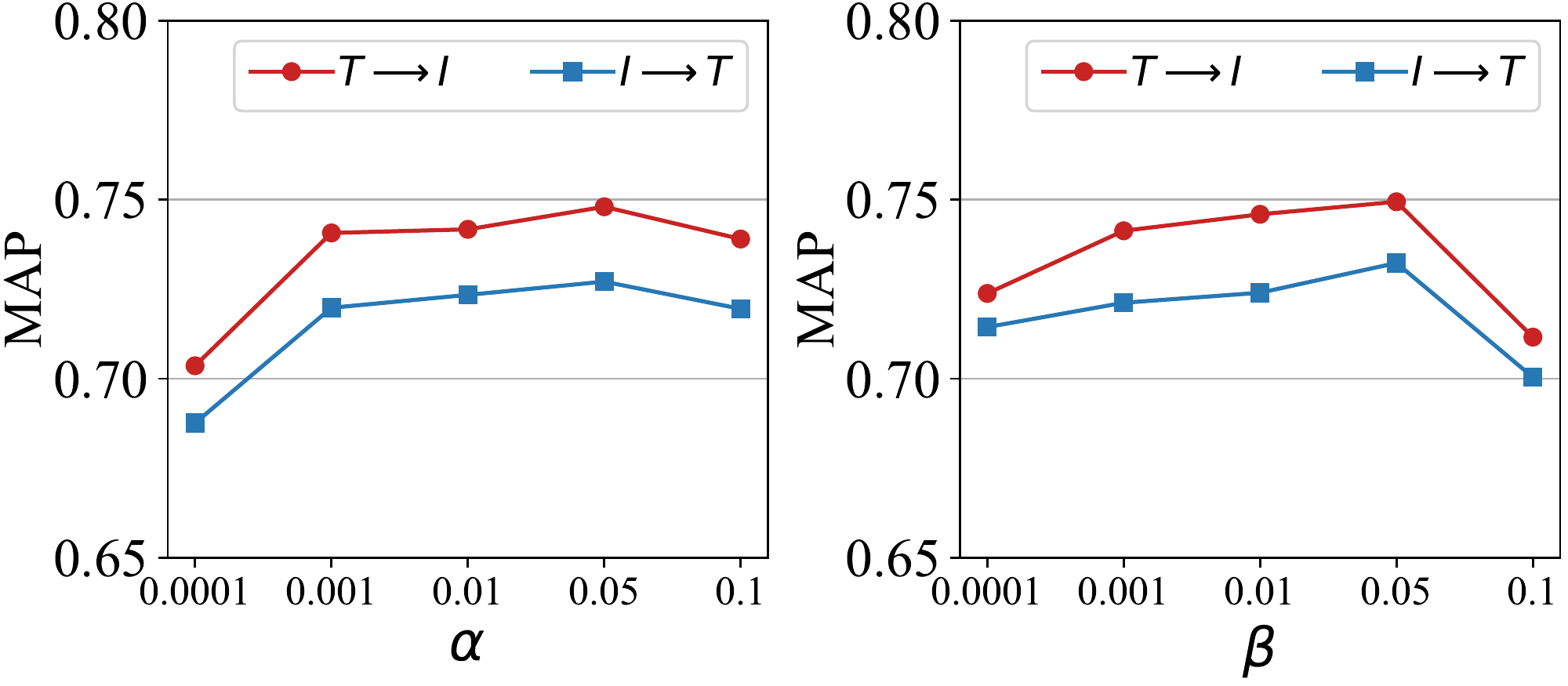}}
\subfigure[NUS-WIDE]{
\includegraphics[width=8cm]{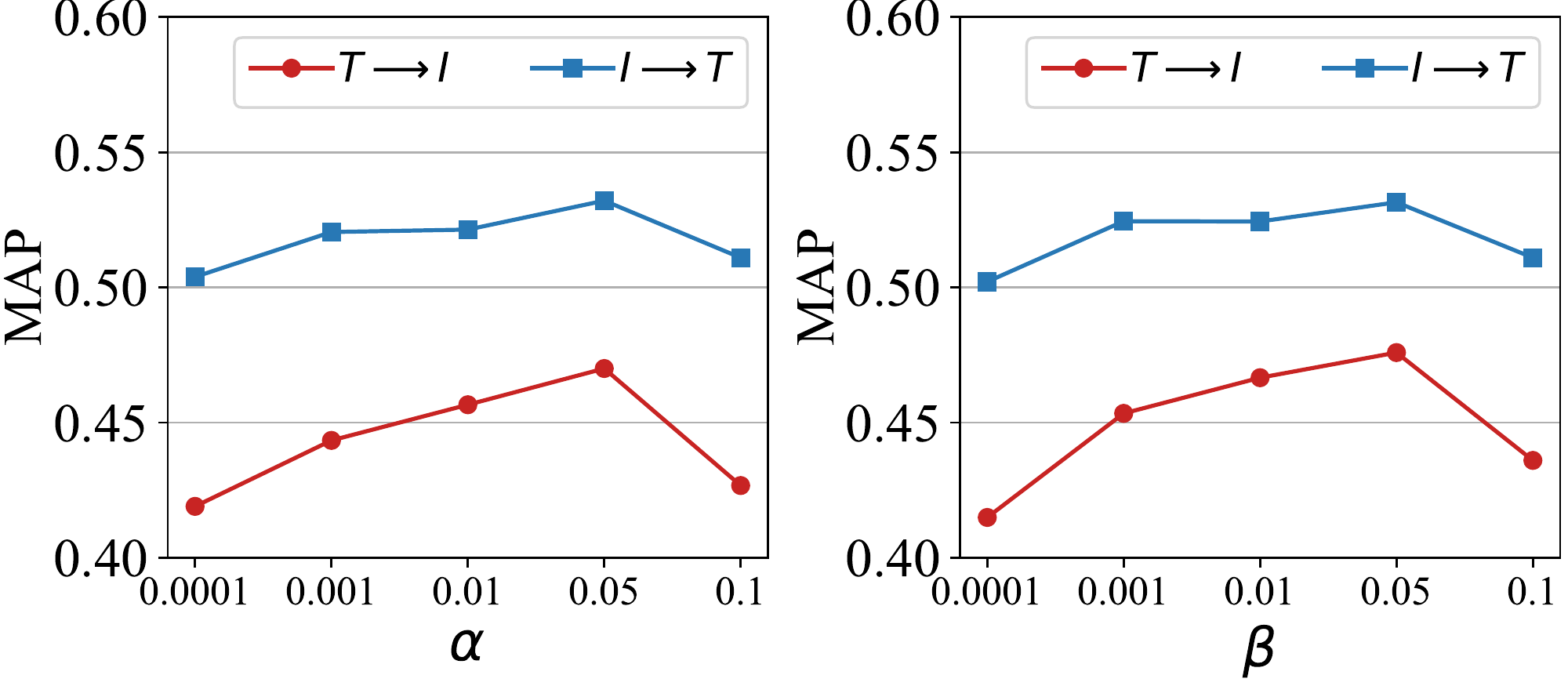}}
\caption{$\alpha$ and $\beta$ vs. MAP on Flicker25K and NUS-WIDE}
\label{fig:alphaBeta}
\end{figure}

\begin{figure}[h!tbp]
\subfigure[Flicker25K]{
\includegraphics[width=8cm]{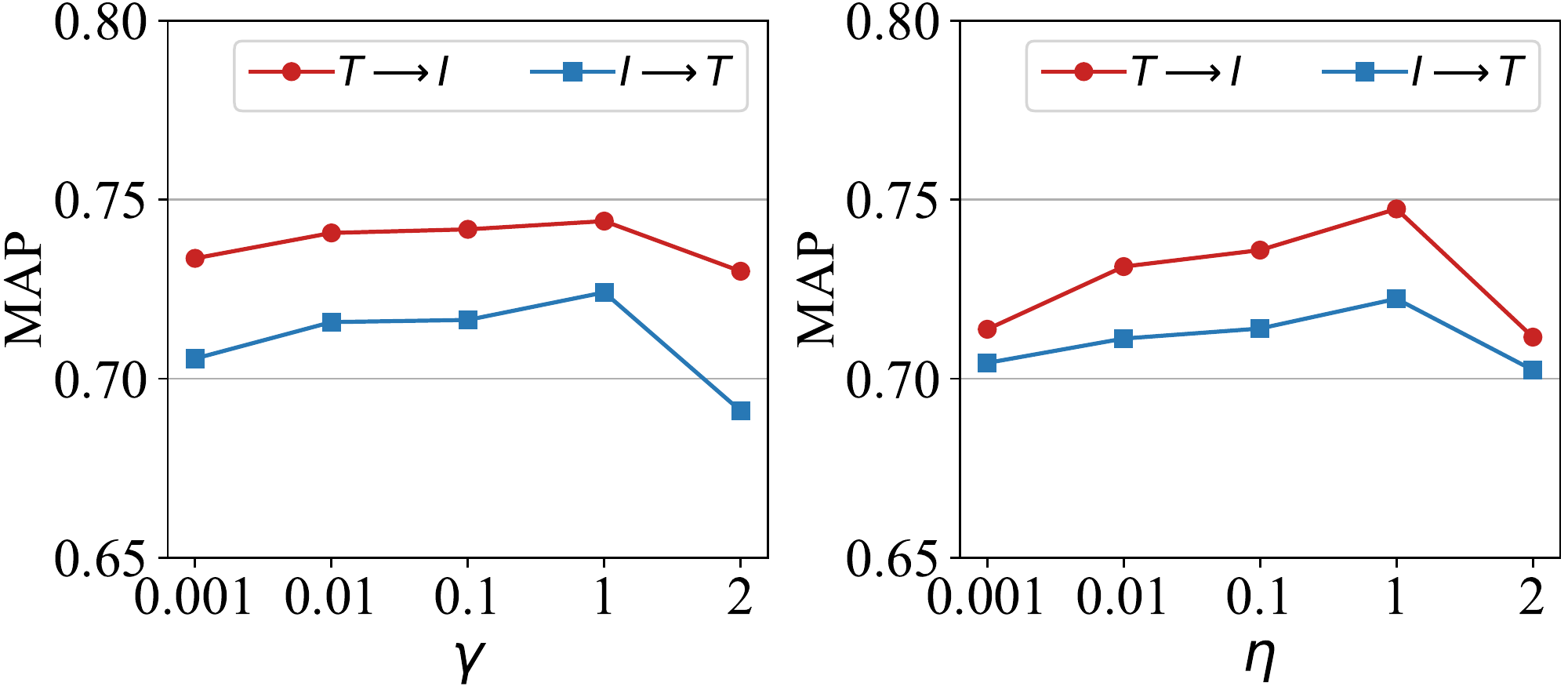}}
\subfigure[NUS-WIDE]{
\includegraphics[width=8cm]{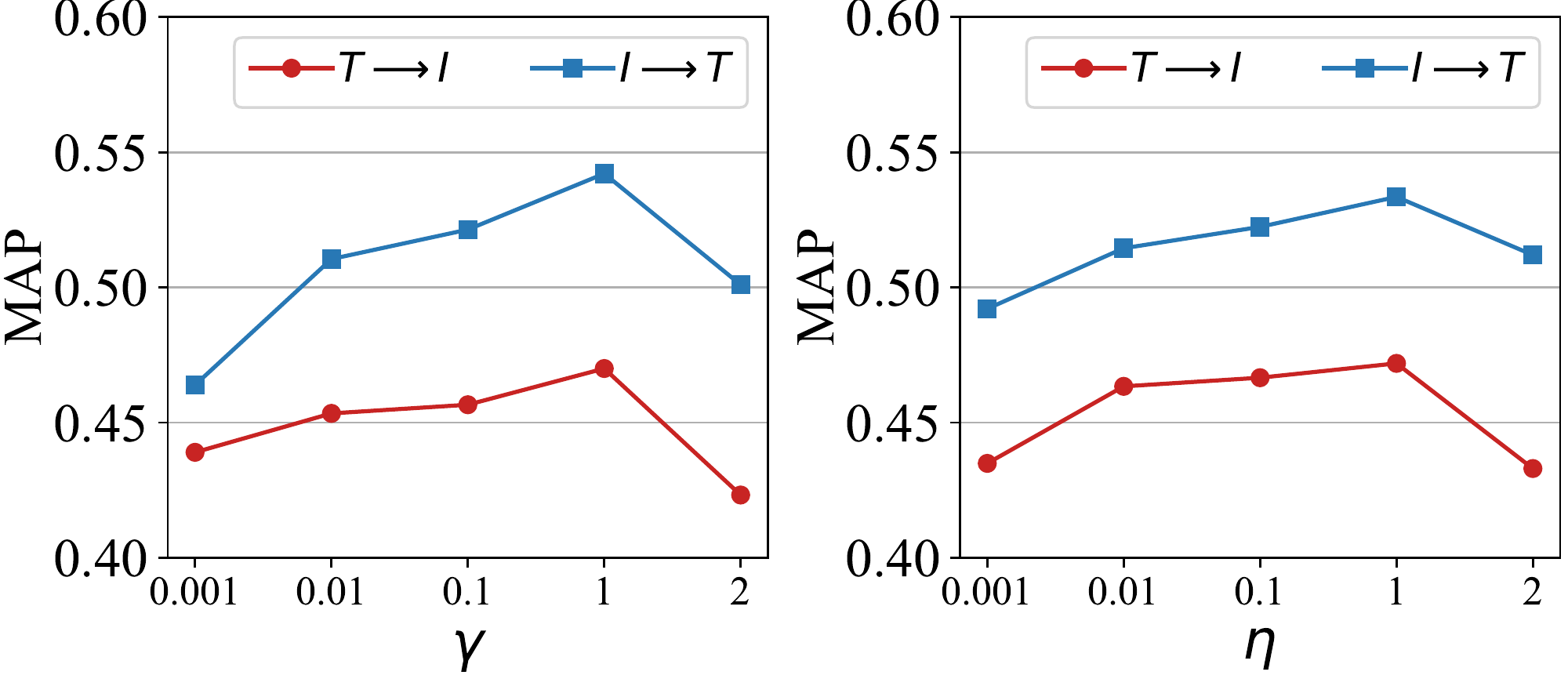}}
 \caption{$\gamma$ and $\eta$ vs. MAP on Flicker25K and NUS-WIDE}
\label{fig:gammaEta}
\end{figure}

Figure \ref{fig:alphaBeta} reveals the MAP results of LtCMH under different input values of $\alpha$/$\beta$ while fixing $\beta = 0.05$/$\alpha = 0.05$ respectively on Flicker25K and NUS-WIDE. Figure \ref{fig:gammaEta} presents the MAP results of LtCMH under different input values of $\gamma$/$\eta$ while fixing $\eta = 1$/$\gamma = 1$ respectively on Flicker25K and NUS-WIDE. 

We find that when $\alpha = 0.05$ or $\beta = 0.05$, the MAP values reach the maximum on two datasets. The stability interval for $\alpha$ and $\beta$ is between 0.001 and 0.05. When $\alpha$ and $\beta$ are smaller than 0.0001 or larger than 0.1, the corresponding performance drops a lot. This is because a too small $\alpha$ weakens the cross-modal regularization, which loses attention to intrinsic distribution and the shared labels of samples within each modality. A too small $\beta$ neglects the contrast and collaboration of individuality from the respective modalities and makes individuality of each modality highly correlated, which cannot enrich tail labels encoded by the individuality. A too large $\alpha$/$\beta$ over-weights the influence of commonality/individuality and makes it hard for AE to reconstruct original features, and drags down the validity of AE. 

To gain an in-depth understanding of $\alpha$ and $\beta$, we study the different impact of them on two datasets. We observe that a too large $\alpha$ or a too small $\beta$ pays more attention to the commonality but neglects the individuality, thus bringing more damage to model performance on NUS-WIDE than Flicker. As  samples from text modality of Flicker are represented as 1386 dimensions BOW vectors while 5018 dimensions in NUS-WIDE, text features of NUS-WIDE are richer  than those of  Flicker so that neglecting the individuality of text modality would lose more information on NUS-WIDE than Flicker. Due to the lack of individuality of text modality on Flicker25K, commonality plays a more important role in enriching  tail labels. As a result, when $\alpha$ is too small or $\beta$ is too large, the performance drops more on Flicker than NUS-WIDE.

The results of the parameter sensitivity analysis experiments for $\gamma$ and $\eta$ on different datasets give similar observations. When $\gamma = 1$ or $\eta = 1$, the MAP values reach the maximum on two datasets. The stability interval of them is between 0.01 and 1. When they are smaller than 0.001 or larger than 2, the performance  drops a lot. Because a too small $\gamma$ will lose the cross-modal similarity in $\mathbf{S}$ preserved by $\mathbf{M}^x$ and $\mathbf{M}^y$, and cannot ensure consistent hashing codes across modalities. A too small $\eta$ cannot make each bit of the hash code balanced on all training samples, which leads to the information provided by each bit not being fully used and wastes the coding space of fixed length hash codes. On the other hand, over-weighting them also drags down the quality of cross-modal hashing functions.

\end{document}